\documentstyle[aps,epsfig,prl]{revtex}
\begin{document}
\title{
"One-dimensional" Coherent States and Oscillation Effects in Metals in a Magnetic
Field}
\author{D.A.Contreras-Solorio\dag, J.A.de la Cruz-Alcaz\dag,
S.T.Pavlov\dag\ddag}
\address{\dag Facultad de Fisica de la UAZ, Apartado Postal C-580,
98060 Zacatecas, Zac., Mexico\\ \ddag P. N. Lebedev Physical Institute, Russian
Academy of Sciences, 119991 Moscow, Russia} \twocolumn[\hsize
\textwidth\columnwidth\hsize \csname @twocolumnfalse
\endcsname
\date{\today}
\maketitle \widetext
\begin{abstract}
\begin{center}
\parbox{6in}
{The "one-dimensional" coherent states are applied to describe  such macroscopic
quantum phenomena as the Shubnikov - de Haas and the de Haas-van Alphen oscillation
effects in metals, semimetals, and degenerate semiconductors.  The oscillatory part
of the electron density of states in a magnetic field is calculated. A substantial
simplification of calculations is achieved.}
\end{center}
\end{abstract}

] \narrowtext

\section{Introduction}
 R. J. Glauber [1,2] introduced in 1963 the concept of a
 coherent state $|\alpha\rangle$ as an eigenstate of a
 non-hermitian annihilation operator $\hat{a}$ of excitations of
 the boson type ($\hat{a}|\alpha\rangle =\alpha |\alpha\rangle$).
 The Schr\"odinger equation for a charge in a constant uniform
 magnetic field reduces to the Schrodinger equation for a
 one-dimensional displaced harmonic oscillator.
 The use of coherent states significantly simplifies
 mathematical calculations of the oscillating part of the
 thermodynamic characteristics. Coherent states are eigenstates
 of a non-hermitian operator and are not orthogonal, i.e. transitions
 between different coherent states can occur spontaneously.
 The Shubnikov-de Haas and de Haas-van Alphen effects are not only
 quantum effects. They are also macroscopic effects, and in these
 respects (the quantum character and macroscopic scale,
 simultaneously) they are related with such phenomena as
 superconductivity, weak-link superconductivity (Josephson
 effects), laser radiation, and von Klitzing's effect (the quantum Hall effect).
 Our aim is to demonstrate clearly ( by using the
 method of coherent states combined with an universal approach to the thermodynamic
 and kinetic effects in
 metals in a constant uniform magnetic field) not only the
 mathematical advantage of such a combination, but also to establish
 the physical reasons for why the mathematical description is
 adequate for the physics of the quantum oscillation effects.
The physical nature of oscillations of the kinetic coefficients
 of a metal in a magnetic field (Shubnikov-de Haas effect) as well
 as oscillations of the thermodynamic potentials and their
 derivatives has been established on the basis of Landau's theory
 of diamagnetism.
 The oscillations are governed by two factors: the presence of the
 Fermi surface and the radical change in the density of states
 $\rho(\varepsilon)$ when the magnetic field is turned on [3].
 Turning on a constant uniform magnetic field $\bf{H}$
 parallel to the z-axis makes the motion of a current-carrying
 particle quasi-one dimensional  and the density of states changes from
 $\rho_{3d}(\varepsilon)\propto\sqrt{\varepsilon}$
 to $\rho_{1d}(\varepsilon)\propto{1\over\sqrt{\varepsilon}}$ (for
 the three- and one-dimensional systems, respectively).Due to the
 Landau quantization of the electron energy spectrum this inverse
 square-root singularity of $\rho(\varepsilon)$ is repeated many
 times  in the energy interval $0\le\varepsilon\le\mu$ ($\mu$ is
 the chemical potential),
 when the condition $\mu\gg\hbar\omega_{H}$ is satisfied (where
 $\omega_{H}=eH/mc$ is the cyclotron frequency; $m, e$ is the
 effective mass and the charge of the current carrier, respectively,
  and $c$ is the light velocity
 in vacuum. For energies $\varepsilon\approx\mu $ near
 the Fermi surface the density of states $\rho (\varepsilon)$ is
 an almost-periodic function of the magnetic field. This is the origin
 for the oscillatory character of the magnetic field dependence of
 both the thermodynamic quantities ("linear" with respect to $\rho
 (\varepsilon)$) and the kinetic coefficients ("quadratic" with respect
 to $\rho(\varepsilon)$). The oscillation  period is the same for
 both types of quantities and is equal to the oscillation period of the
 function $\rho(\varepsilon)$.

\section{Some Thermodynamics Relations}
The thermodynamic potential $\Omega_{H}=F_{H}-\mu N$ is defined by the expression [4]

\begin{equation}
\label{1} \Omega_{H}=-T\sum_{\nu}\ln \left[1+e^{(\mu-\varepsilon_{\nu})/T}\right].
\end{equation}

In the integral form it may be written as

\begin{equation}
\label{2} \Omega_{H}=-T\int_{0}^\infty d\varepsilon\rho(\varepsilon)
\ln\left[1+e^{(\mu-\varepsilon)/T}\right].
\end{equation}

The density of states $\rho(\varepsilon)$ is given by

\begin{equation}
\label{3}\rho(\varepsilon)=\sum_{\nu}\delta(\varepsilon- \varepsilon_{\nu})=Tr
\delta(\varepsilon-\hat{\cal{H}}),
\end{equation}
where $F_{H}$ is the free energy, $N$ is the total number of particles, $T$ is the
temperature (in energy units), $\nu$ is the set of all the quantum numbers
characterizing a single-particle state and $\hat{\cal{H}}$ is the single particle
Hamiltonian. For the thermodynamic potential derivatives we have

\begin{eqnarray}
\label{4}N=-\bigg({\partial\Omega_{H}\over\partial\mu}\bigg)_ {T,V,H},~~~
M=-\bigg({\partial\Omega_{H}\over\partial H}\bigg)_{T,V,\mu},\nonumber\\
C=-T\bigg({\partial^{2}\Omega\over\partial T^{2}}\bigg)_{V,\mu,H},
\end{eqnarray}
where $M$ is the magnetic moment and $C$ is the heat capacity. We shall calculate the
density of states from Eq.(2), setting $T=0$ for simplicity. Then  the
$\rho(\varepsilon)$ transforms into $\rho(\mu)$ and is related with $\Omega_{H}$ by
the expression

\begin{equation}
\label{5}\rho(\mu)=-\bigg({\partial^{2}\Omega_{H}\over\partial\mu^{2}}\bigg)_{V,H,T=0}.
\end{equation}

We can easily see from Eq.(5), the density of states $\rho(\mu)$ at the Fermi
surface  is not only related with the observable quantities presented in Eq.(4). Its
oscillatory part $\tilde{\rho}(\mu)$ contains the period of the oscillations, which
in turn through the Lifshitz-Onsager relation determines the area of the extremal
sections of the Fermi surface by a plane perpendicular to $\bf H $.  The oscillatory
part $\tilde{\rho}(\varepsilon)$ of the density of states also answers the question
about the physical nature of the oscillations of the kinetic coefficients in a
magnetic field. As is well known from the theory of the Shubnikov-de Haas effect, the
appearance of a nonzero current in the direction of the electric field ${\bf E}||{\bf
x}$ is attributable to the appearance of electron scattering, which under the
conditions of the Shubnikov-de Haas effect can be assumed to be elastic [3, 5]. The
fact that $\rho(\varepsilon)$ in Eq.(3) is represented in the form of a trace makes
it possible to employ any complete set of wave functions in the computational
procedure. Oscillatory wave functions (which are eigenfunctions of the number operator
of boson excitation $\hat a^{+}\hat a |n\rangle=n|n\rangle $) do not carry any
information about the presence of the Fermi surface, while for the coherent states
$|\alpha\rangle$ (which are eigenfunctions of the operator $\hat a (\hat
a|\alpha\rangle=\alpha|\alpha\rangle)$) the average number of the particles  is equal
to

\begin{equation}
\label{6} \langle\alpha|\hat{a}^{+}\hat{a}|\alpha\rangle
=\bar{n}_{\alpha}\simeq{\mu\over \hbar\omega_{H}}.
\end{equation}

In addition, the coherent states are characterized by a well-defined phase [2,6]. This
is related with the existence of a phase characteristic(cyclotron period) of the
oscillation phenomena under study. It suggests us to use coherent states for our
problem.

\section{ Coherent states of a charged
 particle in a constant uniform magnetic field}

Coherent states appear when one solves a task for an linear oscillator. Some physical
phenomena (superconductivity, Shubnikov - de Haas, de Haas-van Alphen effects) are
quantum in their physical nature and macroscopical in their scale. Macroscopic scale
indicates on a possibility of an almost classical description of such phenomena. The
coherent states are much more convenient to describe simultaneously a field phase and
amplitude, and to show a connection between the classical and quantum field
description. Historically, L. D. Landau was the first to show that the Schr\"odinger
equation for the eigenfunctions and eigenvalues of a charge in a constant magnetic
field has the form of the Schr\"odinger equation for the one-dimensional linear
oscillator. Coherent states have been used to recast in new terms the theory of
Landau dimagnetism and the theory of the de Haas-van Alphen effects for free electron
gas.

The achievements of the physics of coherent states have not been
extended enough to oscillation effects in  metals with an
arbitrary dispersion relation for electrons or to numerous other
quantum physical phenomena observed in metals in a magnetic field.

We will introduce the coherent states for a charge  in a constant magnetic field
${\bf H} ||{\bf z}$ and the Hamiltonian [7]
\begin{equation}
\label{7}\hat{\cal{H}}={1\over{2m}}\big(\hat{\bf{p}}-{e\over c}\bf{A}\big)
^{2}+\hat{\cal{H}}_{\sigma}=\hat{\cal{H}}_{\bot}+\hat{\cal{H}}_{z}+\hat{\cal{H}}_{\sigma},
\end{equation}

\begin{equation}
\label{8}\hat{\cal{H}}_z={\hat{p}^{2}_z\over 2m},~~~ \hat{\cal{H}}_{\sigma}=-{g\over
2}\mu_{B}\sigma_zH,~~~\sigma_z=\pm1,
\end{equation}
where $\hat{\bf{p}}$ is the momentum operator, $m$ is the  bare electron mass, $g^{*}$
is the effective spectroscopic splitting factor, $\mu_{B}$ is the Bohr magneton.

We choose the vector potential ${\bf A}$ of the magnetic field in the Landau-gauging
[7] as follows

\begin{eqnarray}
\label{9} {\bf A}={\bf A}(-yH, 0, 0),~~~{\bf H}=\nabla\times{\bf A}.
\end{eqnarray}

In this case $\hat{\cal{H}}_{\bot}$ corresponds to an one-dimensional oscillator along
the y axes

\begin{equation}
\label{10} \hat{\cal{H}}_{\bot}={\hat{p}^{2}_{y}\over{2m}}+{1\over{2}}
m\omega^{2}_{H}(y-y_{0})^{2}
\end{equation}
(where  $y_{0}=-cp_{x}/eH$) instead of two coupled oscillators in gauging ${\bf A}=
(1/2)[{\bf H}{\bf r}]$. It gives us a possibility to avoid using of "two-dimensional"
coherent states (see [8]). In dimensionless coordinates

\begin{equation}
\label{11} \eta={y-y_0\over{l}_{H}},~~~
l_{H}=\left({\hbar\over{m}\omega_{H}}\right)^{1/2},
\end{equation}
the Hamiltonian $\hat{\cal{H}}_{\bot}$ takes the form

\begin{equation}
\label{12} \hat{\cal{H}}_{\bot}={1\over{2}}\hbar\omega_{H}(\hat{p}_\eta^2+\eta^{2}),
\end{equation}
$\hat{p}_\eta=-i\nabla_\eta$.We introduce the operators $\hat{a}$ and $\hat{a}^{+}$

\begin{equation}
\label{13}\hat{a}={1\over\sqrt{2}}\left(\eta+{\partial\over \partial\eta}\right),~~
\hat{a}^+={1\over\sqrt{2}}\left(\eta-{\partial\over \partial\eta}\right)
\end{equation}
 $[\hat{a},\hat{a}^{+}]=1$.
Then $\hat{H}_{\bot}$ results in

\begin{equation}
\label{14}\hat{H}_{\bot}=\hbar\omega_{H} \bigg(\hat{a}^{+}\hat{a}+{1\over{2}}\bigg),
\end{equation}
and

\begin{equation}
\label{15}\hat{H}=\hbar\omega_{H}
\bigg(\hat{a}^{+}\hat{a}+{1\over{2}}\bigg)+{\hat{p}^{2}_{z}\over{2m}}
-{g^{*}\over{2}}\mu_{B}\sigma_{B}H.
\end{equation}

Thus, the partial motion of an electron in a magnetic field in the $xy$ plane is
described by the Eq. (14), which contains the operators $\hat{a}$, $\hat{a}^{+}$
(defined in Eq.(13)), satisfying  the Bose commutation relations. With the help of
the operators $\hat{a}$, $\hat{a}^{+}$ we determine the states:

a) the vacuum state $|0\rangle$ such that $\hat{a}|0\rangle=0$;

b) the Fock (after V.A.Fock) state $|n\rangle$, which is an eigenstate of the
operator  $\hat{n}=\hat{a}^{+}\hat{a}$;

\begin{equation}
\label{16} \hat{n}|n\rangle=n|n\rangle,\,\,
|n\rangle={(\hat{a}^{+})^{n}\over\sqrt{n!}}|0\rangle;
\end{equation}

c)the one-dimensional coherent state $|\alpha\rangle$ which is an eigenstate of the
operator $\hat{a}$

\begin{equation}
\label{17}\hat{a}|\alpha\rangle=\alpha| \alpha\rangle.
\end{equation}

The coherent state $|\alpha\rangle$ can be obtained also with the help of the
displacement operator $\hat{D}(\alpha)$

\begin{equation}
\label{18}|\alpha\rangle =\hat{D}(\alpha)|0\rangle,
\end{equation}
where

\begin{equation}
\label{19}\hat{D}(\alpha)=e^{\alpha\hat{a}^{+}-\alpha^{\ast}\hat{a}}= e^{-|
\alpha|^{2}/2}e^{\alpha\hat{a}^{+}}e^{-\alpha^{\ast}\hat{a}}.
\end{equation}

Thus, we have a complete normalized set of wave functions, which are the
eigenfunctions of non-hermitian operator and for this reason are not orthogonal.

It should be specially noted, however, that the partial motion of a fermion
(electron) in the $xy$ plane in the magnetic field $\bf{H}$ is described with the help
of a boson field.

\section{Oscillations of the electron density of states}

We can employ the following complete normalized set of wave functions to calculate
$\rho(\mu)$ of a metal in a quantizing magnetic field

\begin{equation}
\label{20}|\sigma_{z},p_{z};\alpha\rangle= L^{-1/2}_{z}e^{(ip_{z}z/
\hbar)}\chi|\alpha\rangle,
\end{equation}
where

\begin{equation}
\label{21}\hat{\sigma}_{z}\chi=\sigma_{z}\chi,~~~~\sigma_{z}=\pm1,
\end{equation}
$L_{z}$ is the normalization length, and $\hat{\sigma}_{z}$ is the Pauli matrix.

Taking the trace in Eq. (3) and using Eq. (20), we obtain
\begin{eqnarray}
\label{22} \rho(\mu)=\sum_{p_{z},\sigma_{z}}\int
{d^{2}\alpha\over\pi}\langle\alpha;P_{z},\sigma_{z}|\delta(\mu-H) |\sigma_{z},p_{z};
\alpha\rangle
\nonumber\\
 ={L_{z}\over\pi(2\pi\hbar)^{2}}
\sum_{\sigma_{z}}\int_{-\infty}^{\infty} dp_{z}\int
d^{2}\alpha\int_{-\infty}^{\infty}dt\nonumber\\
\times\langle\alpha;p_{z}, \sigma_{z}| e^{i(\mu-\hat{H})t/\hbar}|
\sigma_{z},p_{z};\alpha\rangle,
\end{eqnarray}
where $d^{2}\alpha=d(Re\,\alpha)d(Im\,\alpha)$. In the operator $\hat{H}$ all three
terms commute one with another. We obtain the following relations:

\begin{equation}
\label{23} \sum_{\sigma_{z=\pm1}}e^{i(t/\hbar)(g^{*}/2)
\mu_{B}H}=2\cos\left({g^{*}\mu_{B}H\over2\hbar}t\right);
\end{equation}

\begin{equation}
\label{24}\int_{-\infty}^{\infty}dp_{z}e^{-{itp_{z}^{2}\over 2m\hbar}}
=\left({2\pi\hbar m\over|t|}\right)^{1/2}e^{-i{\pi\over4}sign\,t};
\end{equation}

\begin{eqnarray}
\label{25} \langle\alpha\mid e^{-it\omega_{H}\hat{a}^{+}\hat{a}}|\alpha\rangle
=\sum_{n=0}^{\infty}\langle\alpha\mid e^{-it\omega_{H}\hat{a}^{+}\hat{a}}|
n\rangle\langle n| \alpha\rangle\nonumber\\ =\sum_{n=0}^{\infty}e^{-it\omega_{H}n}|
\langle n| \alpha\rangle |^{2} =e^{-|\alpha|^{2}\left(1-e^{-it\omega_{H}}\right)}.
\end{eqnarray}

In the last equation we have used the condition

\begin{equation}
\label{26}\sum_{n=0}^{\infty}|n\rangle\langle n|=1
\end{equation}
of the completeness of the set of the Fock states.The result for the scalar product of
the Fock and coherent states is as follows

\begin{eqnarray}
\label{27}\langle n|\alpha\rangle=\left\langle n\left|e^{-|\alpha|^{2}/2}
e^{\alpha\hat{a}^{+}}\right|0\right\rangle\nonumber\\
={\alpha^{n}\over\sqrt{n!}} e^{-|\alpha|^{2}/2}.
\end{eqnarray}

The density of states $\rho(\mu)$ at the Fermi surface results in the form of a
single integral

\begin{eqnarray}
\label{28} \rho(\mu)&=&{ L_{Z}\Phi
m^{1/2}\over(2\pi\hbar)^{3/2}\Phi_0}\int_{-\infty}^{\infty} dt{e^{i({\mu
t\over\hbar}-{\pi\over 4}sign\,t)} \cos{\big({g^{*}\mu_{B}H\over 2\hbar}t\big)}\over
i|t|^{1/2}\sin{(t\omega_{l})}},\nonumber\\
\Phi&=&L_xL_yH,~~~\Phi_0={ch\over e}.
\end{eqnarray}

It is calculated with the help of the residue theorem by integrating along the contour
shown in Fig. 1.

The oscillating part of the density of states $\tilde{\rho}(\mu)$ is determined by
the contribution into the integral of the poles located on the real axis at the points

\begin{eqnarray}
\label{29} t_{k}={2\pi\over\omega_{H}}K,~~~K=\pm1,\pm2,\pm3,...
\end{eqnarray}
and has the form

\begin{eqnarray}
\label{30}\tilde{\rho}(\mu)={mV\over\pi^2\hbar^{2}}
\left({eH\over c\hbar}\right)^{1/2}\sum_{K=1}^{\infty}K^{-1/2}\nonumber\\
\times\cos{\left({\pi g^{*}m\over 2m_{0}}K\right)} \cos{\left(2\pi
K{\mu\over\hbar\omega_{H}}-{\pi\over 4}\right)},
\end{eqnarray}
which contains a period of the oscillations

\begin{equation}
\label{31} \Delta\Big({1\over H}\Big)={e\hbar\over mc\mu}.
\end{equation}

\section{Discussion}
Our approach to the analysis of oscillation effects in a metal in a magnetic field
makes it possible not only to substantially simplify the mathematical procedure as
compared with the traditional method of analyzing these phenomena, but also to study
in a universal manner both the thermodynamics and kinetic effects, to easily extend
the analysis to the case of current carriers with an arbitrary energy spectrum and
nonzero temperature, and to include the effect of scattering on the form of the
oscillation dependence.

The physical reason for the simplification achieved in the mathematical procedure is
that  the coherent state employed in the calculations  describe quantum macroscopic
and therefore also quasi-classical phenomena, which are the Shubnikov-de Haas and de
Haas-van Alphen effects in metals, semimetals and degenerate semiconductors.

\section{Acknowledgements}
        S.T.P thanks the Zacatecas Autonomous University and the National
Council of Science and Technology (CONACyT) of Mexico for the financial support and
hospitality. D.A.C.S. thanks CONACyT (27736-E) for the financial support.

%%%%%%%%%%%%%%%%%%%%%%%%%%%%%%%%%%%%%%%%%%%%%%%%%%%%%%%%
%%%%%%%%%%%%%%%%%%%%%%%%%%%%%%%%%%%%%%%%%%%%%%%%%%%%%%%%
\begin{figure}
\caption {The integration contour in the complex plane $t$ for the calculation of the
integral Eq. (28). }
\end{figure}

\end{document}